\documentclass[manuscript]{aastex61}
\usepackage{txfonts}
\usepackage{latexsym,bm}
\usepackage{multirow}
\newcommand{\dee}{\mathrm{d}}

\shorttitle{perpendicular shock acceleration}

\shortauthors{Kong et al.}

\begin{document}

\title{NUMERICAL SIMULATIONS OF PARTICLE ACCELERATION AT
INTERPLANETARY QUASI-PERPENDICULAR SHOCKS}

\correspondingauthor{G. Qin}
\email{qingang@hit.edu.cn}

\author{F.-J. Kong}
\affiliation{State Key Laboratory of Space Weather, National
Space Science Center, Chinese Academy of Sciences,
Beijing 100190, China}
\affiliation{College of Earth Sciences, University of
Chinese Academy of Sciences, Beijing 100049, China}
\author[0000-0002-3437-3716]{G. Qin}
\affiliation{School of Science, Harbin Institute of Technology, Shenzhen, 
518055, China}
\author{L.-H. Zhang}
\affiliation{State Key Laboratory of Space Weather, National
Space Science Center, Chinese Academy of Sciences,
Beijing 100190, China}
\affiliation{National Astronomical Observatories, Chinese Academy of Sciences; 
Key Laboratory of Solar Activity, Chinese Academy of Sciences}

\begin{abstract}
Using test particle simulations we study particle acceleration at highly 
perpendicular ($\theta_{Bn}\geq 75^\circ$) shocks under conditions of modeling
magnetic turbulence. We adopt a backward-in-time method to solve the 
Newton-Lorentz equation using the observed shock parameters for 
quasi-perpendicular interplanetary shocks, and compare the simulation results 
with $ACE$/EPAM observations to obtain the injection energy and timescale of 
particle acceleration. With our modeling and observations we find that a large 
upstream speed is responsible for efficient particle acceleration. Our results 
also show that the quasi-perpendicular shocks are capable 
of accelerating thermal particles to high energies of the order of MeV for both 
kappa and Maxwellian upstream distributions, which may originate from 
the fact that in our model the local background magnetic field has a 
component parallel to the shock normal.
\end{abstract}


\section{INTRODUCTION}
Many energetic particle events in space are associated with the acceleration at 
collisionless shocks \citep[e.g.,][]{Heras1992, Kallenrode1997, Zank2000, Li2003, 
Rice2003, WangEA12, QinEA13, QiEA17}. 
The most popular theory for charged-particle acceleration is diffusive shock 
acceleration (DSA) \citep{Krymsky1977DoSSR...234..1306K,Axford1977ICRC...11..132A,
Bell1978MNRAS...182..147B,Blandford1978ApJ...221..L29B},
which is capable to predict a power-law distribution downstream of the shock with
a spectral exponent that depends only on the density compression ratio.
It is widely accepted that this theory explains many important energetic particle
events related to collisionless shocks. However, the energy spectra observed in 
space are not always in agreement with the prediction of DSA theory. In some cases, 
the observed energy spectra have exponential-like rollovers at higher energies
\citep{Ellison1985ApJ...298..400E}. The reason for the spectra rollovers is related 
to the losses of particles, limited acceleration time, or complicated shock 
geometries, all of which are not considered in the basic diffusive theory. 

The angle between the shock normal and the upstream magnetic field, $\theta_{Bn}$,
is one of the most important effects that control the injection of 
particles at shocks. For quasi-parallel shocks, thermal particles can move along 
the magnetic field lines easily to cross the shock front repeatedly, so that they 
can be accelerated to higher energies 
\citep{Ellison1981GRL...8..991E,Quest1988JGR...93..9649Q,
Scholer1990GRL...17..1821S,Giacalone1992GRL...19..433G}.
In contrast, the acceleration of thermal particles at quasi-perpendicular shocks 
is more difficult to be understood because particles tied to the magnetic field 
lines nearly parallel to the shocks can be convected across the shock to downstream. 
Many theoretical works have been devoted to the well-known injection problem at 
quasi-perpendicular shocks. In earlier studies, \citet{Ellison1995ApJ...453..873E} 
showed that thermal particles could be accelerated by quasi-perpendicular shocks, 
but the acceleration efficiency decreases with increasing shock-normal angle. Using
hybrid simulations, \citet{Giacalone2000JGR...105..12541G} and
\citet{Giacalone2003PSS...51..659G} also found that quasi-perpendicular shocks 
did not accelerate particles efficiently. Subsequently, 
\citet{Giacalone2005ApJ...628..L37G} performed large-scale self-consistent plasma 
simulations and demonstrated that thermal particles could be accelerated by 
perpendicular shocks. It is assumed that with large-scale magnetic 
fluctuations, the particles tied to meandering magnetic field lines can experience 
multiple crossings of the shock front, and thus be diffusively accelerated. 

Recently, the injection mechanisms of shock acceleration are investigated. The
injection energy can be derived from diffusion models assuming small anisotropy 
in the particle distribution \citep{Giacalone1999ApJ...520..204G, 
Giacalone2006SSR...124..277G,Zank2006JGR...111..A06108Z}. 
\citet{NeergaardParker2012ApJ...757..97P} and \citet{NeergaardParker2014ApJ...782..52P}
explored the injection energy at quasi-parallel and quasi-perpendicular shocks, 
respectively, through solving the steady-state transport equation. For quasi-parallel 
shocks, it is suggested that the injection energy is small so an upstream Maxwellian 
distribution is sufficient to provide the particle injection for DSA 
\citep{NeergaardParker2012ApJ...757..97P}. The Maxwellian distribution has the form
\begin{equation}
f_0(\boldsymbol p)=n_1\left(\frac{1}{2\pi mkT}\right)^{3/2}\exp
\left(-\frac{(\boldsymbol p-\boldsymbol p_0)^2}{2mkT}\right),
\label{eq:maxwellian}
\end{equation}
where $n_1$ is the upstream particle (solar wind proton) number density, $m$ is 
the mass of a proton, $k$ is Boltzmann's constant, $T$ is the temperature of the 
distribution, $p$ is the proton momentum, and $p_0$ corresponds to the solar wind 
momentum in the spacecraft frame. But for quasi-perpendicular shocks the injection 
energy is assumed to be large. In addition, it is known for quasi-perpendicular 
shocks that all reflected ions can return to the shock front and form a suprathermal 
population in the upstream region, and that a kappa distribution with an enhanced
tail of energetic particles is demonstrated to provide an appropriate particle 
injection for DSA \citep{NeergaardParker2014ApJ...782..52P}.
The kappa distribution can be written as
\begin{equation}
f_0(\boldsymbol p)=\frac{n_1}{[\pi(2\kappa-3)mkT]^{3/2}}
\frac{\Gamma(\kappa+1)}{\Gamma(\kappa-1/2)}\left[1+
\frac{(\boldsymbol p-\boldsymbol p_0)^2}{(2\kappa-3)mkT}\right]^{-\kappa-1}.
\label{eq:kappa}
\end{equation}

In this paper, we perform test particle simulations of particle acceleration 
associated with quasi-perpendicular interplanetary shocks with modeling magnetic 
turbulence by numerically solving the Newton-Lorentz equation with a 
backward-in-time method to clarify the crucial shock features that are responsible 
for efficient particle acceleration. In Section 2, we describe the shock model and 
numerical methods. The numerical results are shown in Section 3. We present in 
Section 4 the conclusions and discussion.

\section{MODEL}
\begin{figure}
\epsscale{0.5} \plotone{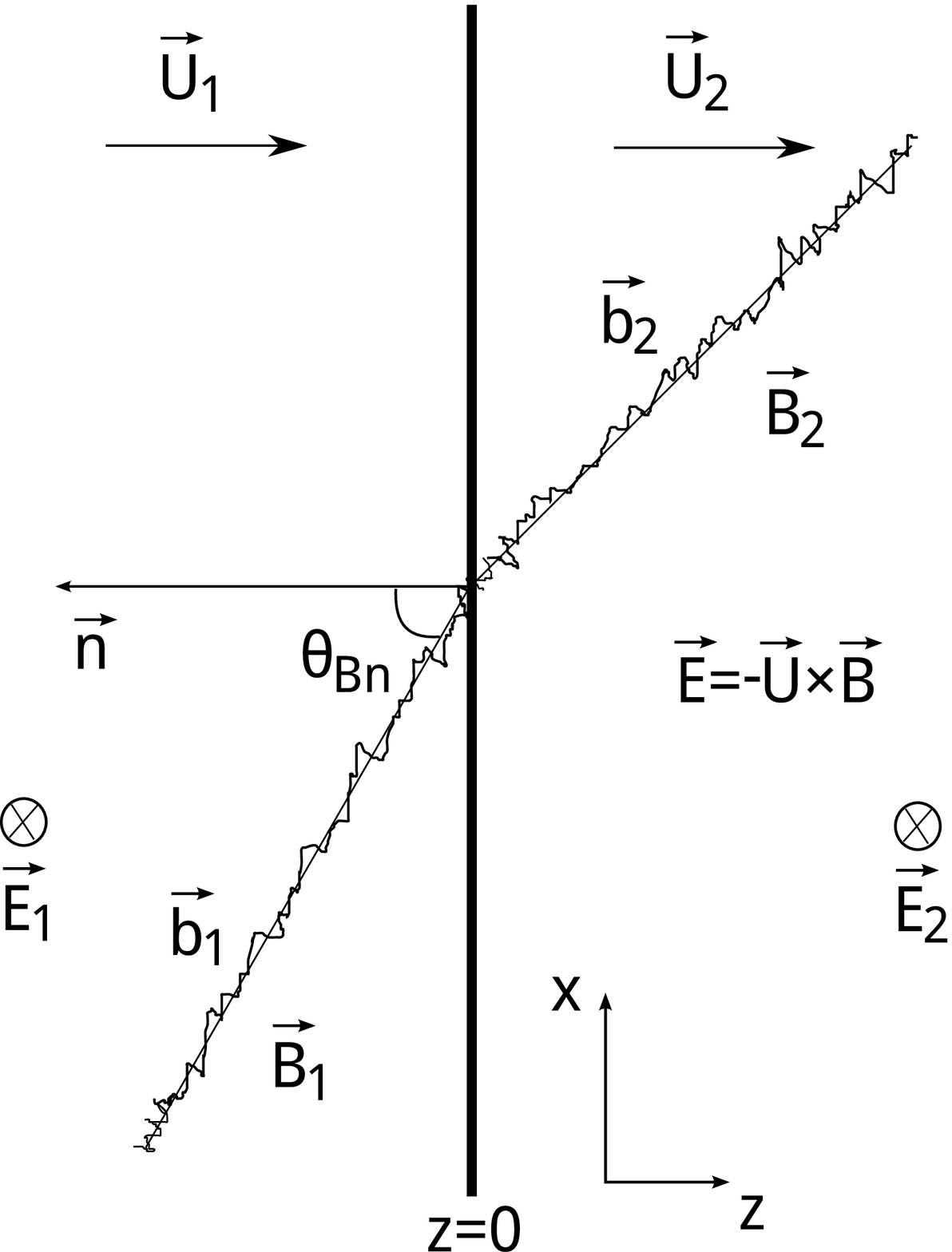} \caption{Geometry of the
shock in our numerical simulations.\label{shock}}
\end{figure}

Our numerical simulations, performed by using the code from \citet{ZhangEA17} to 
solve the Newton-Lorentz equation to obtain the trajectories of particles 
in the shock frame of reference, are similar to those of 
\citet{Decker1986ApJ, Decker1986JGR...91..13349D} except that we here consider a 
two-component magnetic turbulence model (described below). The Newton-Lorentz 
equation for a test particle in the electric field $\boldsymbol{E}$ and 
magnetic field $\boldsymbol{B}$ is given by
\begin{equation}
 \frac{\dee\boldsymbol{p}}{\dee t}=q\left[\boldsymbol{E}(\boldsymbol{r},t)
+\boldsymbol{v}\times\boldsymbol{B}(\boldsymbol{r},t)\right],
\end{equation}
where $\boldsymbol{p}$ is the particle momentum, $\boldsymbol{v}$ is the particle 
velocity, $q$ is the particle charge, $t$ is time, and the frame of reference is 
moving with the shock front. The convective electric field is
$\boldsymbol{E}=-\boldsymbol{U}\times\boldsymbol{B}$, with $\boldsymbol{U}$ the 
plasma flow speed. We consider a planar, fast-mode, collisionless shock with the 
geometry illustrated in Figure \ref{shock}. The shock is located at $z=0$, and 
plasma flows in the positive $z$-direction with the speeds $U_1$ and $U_2$ in the 
upstream and downstream regions as measured in the shock frame of reference,
respectively. In the shock transition the plasma speed is given by
\begin{equation}
U(z)=\frac{U_1}{2s}\left\{(s+1)+(s-1)\tanh\left[\tan\left(-\pi z/L_{diff}
\right)\right]\right\},
\end{equation}
here, $L_{diff}$ is the thickness of the shock transition which is assumed to be 
small enough compared to the gyro-radii of particles in the magnetic field, and 
$s=U_1/U_2$ is the compression ratio. The magnetic field is taken to be time 
independent as
\begin{equation}
\boldsymbol{B}\left(x^\prime,y^\prime,z^\prime\right)=\boldsymbol{B_0}+\boldsymbol{b}
\left(x^\prime,y^\prime,z^\prime\right),
\end{equation}
where $\boldsymbol{B_0}$ is the uniform background field,
$\boldsymbol{b}$ is a zero-mean turbulent field being 
transverse to $\boldsymbol B_0$, and a Cartesian coordinate 
$\left(x^\prime,y^\prime,z^\prime\right)$ system is adopted with $z^\prime$-axis 
to be parallel to the mean magnetic field $\bm{B}_0$. The turbulent magnetic 
fluctuations $\bm{b}$ are composed of a slab component and a 2D one
\citep{
	Mattaeus1990JGR...95..20673M,
	Zank1992JGR...97..17189Z,
Bieber1996JGR...101..2511B,Gray1996GRL...23..965G,
Zank2006JGR...111..A06108Z},
\begin{equation}
\boldsymbol{b}\left(x',y',z'\right)=\boldsymbol{b}^{slab}\left(z'\right)+
\boldsymbol{b}^{2D}\left(x',y'\right),
\end{equation}
where both the two components are perpendicular to the large-scale
background magnetic field $\bm{B}_0$, and have Kolmogorov's spectra with a power 
index $\nu=-5/3$ at high wave-number $k$. The slab component is created with 
fast Fourier transform (FFT) in a box with size 
$L_{z^\prime}=1000\lambda_{slab}$ along the mean magnetic field $\bm{B}_0$ and 
number of grids $N_{z^\prime}=2^{21}=2097152$, where $\lambda_{slab}$ is the 
correlation scale of the slab component and is set to be $0.02$ AU at $1$ AU.
The 2D component is created with FFT in a two-dimensional box with the correlation 
scale length $\lambda_{2D}=0.1\lambda_{slab}$, the 2D box size 
$L_{x^\prime}=L_{y^\prime}=100\lambda_{slab}$, and the number of grids 
$N_{x^\prime}=N_{y^\prime}=4096$. 
We take the ratio of magnetic energy densities of different components in the 
turbulence to be $E_{slab}:E_{2D}=20:80$. For more details on the slab-2D 
magnetic field model, see \citet{QinEA02GRL, QinEA02APJ}, 
\citet{Qin2002PhDT...1Q},
and \citet{ZhangEA17}. Note that the bulk velocity $\boldsymbol{U}$ and mean 
magnetic field $\bm{B}_{0}$ satisfy the Rankine-Hugoniot (RH) conditions, but 
we do not consider RH for the magnetic turbulence for simplicity based on the 
small level and zero average of turbulence.

In the diffusive shock acceleration model, energetic particles can gain energy 
by crossing the shock back and forth because of the magnetic field turbulence 
scatterings. In this work we adopt the time-backward method to calculate 
the trajectories of a large number ($N_0$) of test particles, which are 
released with the same target energy $E_t$ measured in the spacecraft frame at 
the shock front. Note that $N_0$ varies from $6000$ to $25000$ in different events. 
Each particle orbit is followed with numerical calculations by a {\it fourth-order Runge-Kutta}
method with an adjustable time step which maintains accuracy of the order of $10^{-9}$ 
until the test particle's motion time is up to the preset time boundary $t_b$. 
At the time boundary, the particles have different energy $E_b$ (with the 
corresponding momentum $p_b$), most of which are lower than the target energy 
$E_t$ with the time-backward calculation because of shock acceleration. 
In addition, a few number ($N^\prime$) of particles with $E_b$ lower than a preset 
energy $E_{b0}$ which is larger than that associated with the background plasma speed, 
are abandoned. Since $N^\prime$ is much smaller than $N_0$, the fact that we 
abandon the $N^\prime$ particles does not affect the simulation results. 
Note that our simulation box is large enough to ensure particles not to escape 
the acceleration region, so the spatial boundary is not taken into account.
We consider eight target energies according to the energy channels in the range 
(47--4750) keV from the Low-Energy Magnetic Spectrometer 30 (LEMS30) and LEMS120 
detectors in the Electron, Proton, and Alpha Monitor (EPAM) instruments onboard 
the $\mathit{Advanced\ Composition\ Explorer\ (ACE)}$ spacecraft. For each of 
the $N_E=8$ target energies $E_{i}$ we consider target pitch angle cosine with 
an uniform distribution in the spacecraft frame, so the distribution function 
$f_a\left(p_i\right)$ at the shock front is obtained from an initial upstream 
background distribution function $f_0\left(p\right)$ through
\begin{equation}
f_a\left(p_i\right)=\frac{1}
{N_{i}}\sum\limits^{N_{i}}_{j=1}f_0\left(p_{bij}\right),
\end{equation}
where $i=1,\dots,N_E$ identifies the channels of energy, $p_{bij}$ indicates the
boundary momentum of the $j^{th}$ test particle, and $N_{i}=N_0-N_{i}^\prime$ 
denotes the number of injected particles that contribute to the statistics. 
The corresponding accelerated energy spectrum is given by
\begin{equation}
j\left(E_i\right)=f\left(p_i\right)p_i^2.
\end{equation}
For the upstream background distribution of energetic particles we choose the 
kappa distribution as shown in Equation (\ref{eq:kappa}), and we also choose the
Maxwellian distribution as shown in Equation (\ref{eq:maxwellian}) for the comparison.

In our work for each of the shock events with the preset energy $E_{b0}$, 
we gradually increase the preset simulation time until the accelerated particle 
energy spectra do not change obviously, and we thus obtain steady-state spectra 
and the acceleration time $t_{acc}$. We then obtain the injection energy $E_{inj}$ 
from the best $E_{b0}$ to fit steady-state spectra from simulations with EPAM 
observations. The process for the fit is done by examining the dependence 
of the target energy spectrum on the injection energy, and in our simulaitons the 
injection energy error is up to the adopted steps of 0.1 keV and 0.01 keV for 
kappa and Maxwellian distributions, respectively.

Our model adopts test particle method, so energetic particles do not affect shock 
waves and magnetic fields. In addition, We adopt a particle-splitting algorithm 
to improve the statistics of the accelerated particles. This method has been used 
in previous studies by \citet{Giacalone1992GRL...19..433G}. In our present
time-backward simulations, when the energy of a particle drops below a preset 
threshold (0.75 of the initial energy), two new particles are created at the 
location of the mother particle with slight velocity vector deviation. The weight 
of the split particles is one half of the weight of the mother particle prior to the split.

Next, we describe the turbulence level settings. In
\citet{NeergaardParker2014ApJ...782..52P} the steady-state diffusion transport 
equation was solved, and in \citet{Giacalone2015ApJ...799..80G} the time-dependent 
transport equation was solved by  using a diffusion coefficient corresponding to 
a turbulence level which is significantly larger than that observed. 
There is a general idea that the turbulence would be enhanced when the 
position is close to the shock front. In this work, we employ a decaying
turbulence level model from the shock front which has the form
\begin{equation}
b/B_0 = \left\{
\begin{array}{rl}
c\left(\frac{1}{b_1-az}+d_1\right), &  z<0\\
c\left(\frac{1}{b_2+az}+d_2\right), &  z \ge0
\end{array} \right.
\label{eq:decayed}
\end{equation}
where $a$, $b_1$, $d_1$, $b_2$ are the parameters assumed according to 
magnetic turbulence variations, $c$ is the fitting parameter varying with 
different events, and $d_2=d_1+1/b_1-1/b_2$ to make $b/B_0$ continuous at $z=0$. 
For simplicity, we assume for different events that parameters   
$a$, $b_1$, $d_1$, $b_2$, and $d_2$ do not change, with $c$ being the only 
variable. We use $16$ s averaged observation data of $b/B_0$ as a function of 
time $t$ from the Magnetometer Instrument (MAG) onboard $\mathit{ACE}$ preceding 
and following the shock arrival during different events, where positive and 
negative $t$ indicate the time after and before the shock front crossing, 
respectively. Here we assume $b/B_0$ as a function of $z$ which is the distance 
to the shock front with $z=V_{sh}t$, where $V_{sh}$ is the shock speed in the 
spacecraft frame. 

\section{DATA}
Because the main ingredient in the accelerated ions from EPAM observations 
are protons, we only consider proton acceleration at shocks. We select SEP events 
observed at 1 AU during 1998-2005 at quasi-perpendicular shocks with 
$\theta_{Bn}\ge75^\circ$ that are identified from the $\ ACE$ shock lists provided 
by the ACE Web site (\url{http://www.ssg.sr.unh.edu/mag/ace/ACElists/obs_list.html}).
In this work we only include quasi-perpendicular forward shocks where there was no 
additional shock within the previous 20 hours by following 
\citet{NeergaardParker2014ApJ...782..52P}.  

Firstly, similar to \cite{NeergaardParker2014ApJ...782..52P}, we construct the 
upstream kappa particle distribution with $\kappa=4$ using density, velocity, 
and temperature values obtained by averaging the five data points from the 64 s 
averaged data from the Solar Wind Electron, Proton, Alpha Monitor (SWEPAM) 
instrument onboard $\mathit{ACE}$ ahead of the shock arrival, as listed in Table 
\ref{inputpara1}. Secondly, we construct a Maxwellian distribution as shown in 
Equation (\ref{eq:maxwellian}) with density, velocity, and temperature derived 
from observed plasma parameters for modeling the upstream particle distribution.

\begin{figure}
\epsscale{1.} \plotone{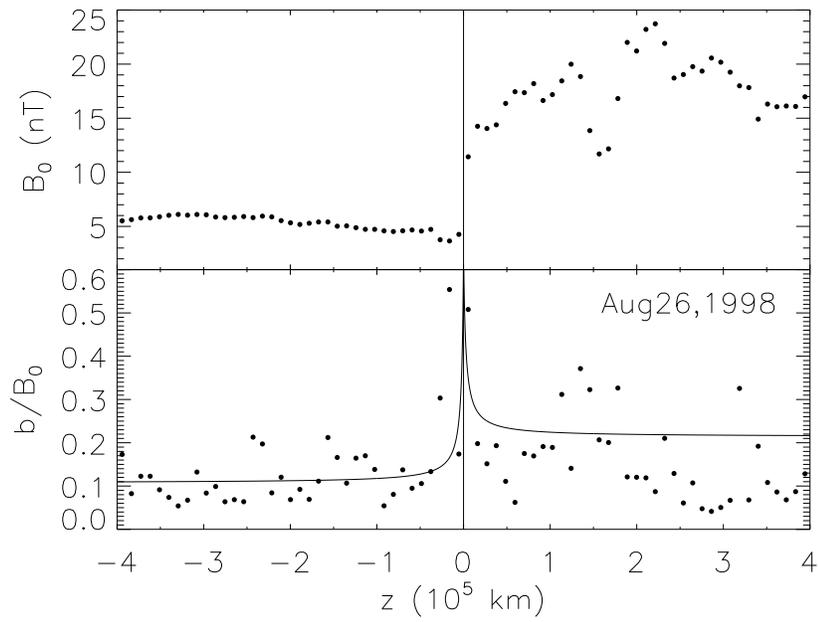} \caption{The
$ACE$/MAG observations of the magnetic field (top panel) and the ratio of the rms
value to the total magnetic field (bottom panel) are plotted with solid circles during
the shock event on 1998 August 26. The time of the shock passage is indicated by a vertical 
black line. We also plot in the bottom panel
the constructed turbulence model for $b/B_0$ with a solid line.\label{bdb}}
\end{figure}

Comparing MAG data onboard $\mathit{ACE}$ for different shock events and the 
decaying turbulence level model with Equation (\ref{eq:decayed}), we get the 
parameters $a=0.0001$ km$^{-1},\ b_1=0.2,\ d_1=1, \ b_2=0.25$, and $d_2=2$. 
In addition, we get $c$ for each event. In Figure \ref{bdb}, we plot the 
$ACE$/MAG observations of the magnetic field and the ratio of the 
root mean square (rms) value to the total magnetic field during a 
representative shock event on 1998 August 26. It is shown that there is an 
abrupt jump of $b/B_0$ at the time of the shock passage. Besides the data from 
spacecraft, we also plot the results from our model with Equation (\ref{eq:decayed}).
Table \ref{inputpara2} summarizes the input parameters of the shock events selected,
such as the shock obliquity $\theta_{Bn}$, shock speed $V_{sh}$, upstream speed $U_1$,
Mach number $M_A$, compression ratio $s$, upstream magnetic field strength $B_0$, the
turbulence parameter $c$, and the upstream convective electric field. The upstream
background magnetic field is from $320$ seconds averaged data of MAG onboard $ACE$.

\section{SIMULATION RESULTS}

\begin{figure}
\epsscale{1.} \plotone{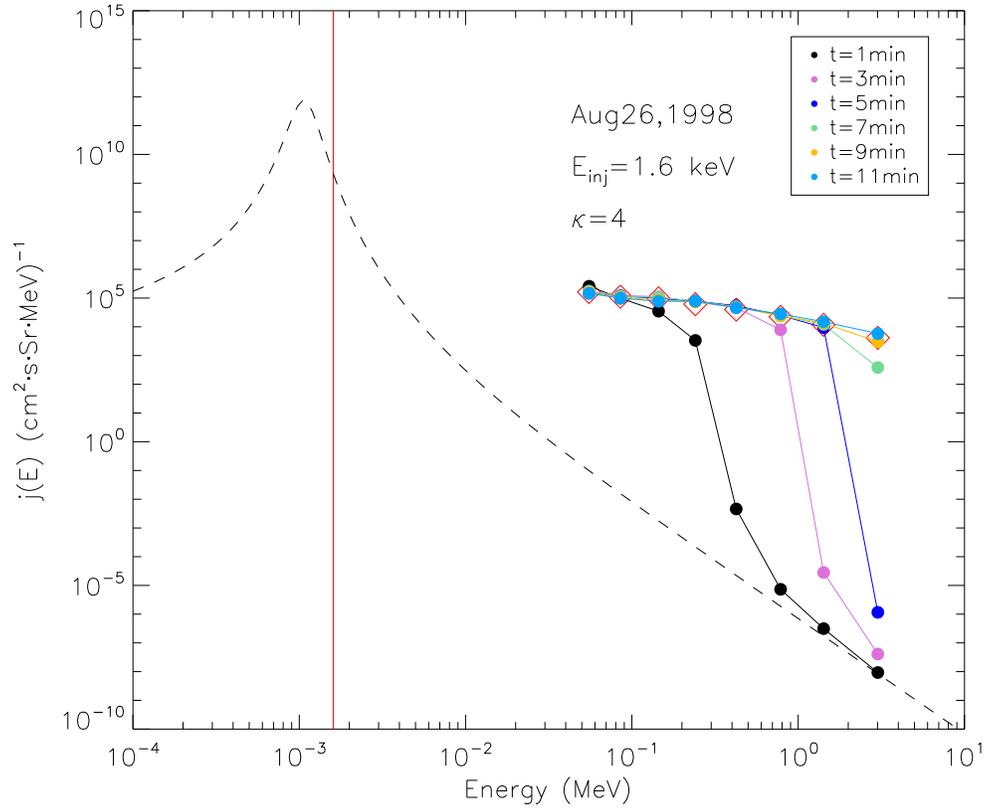} \caption{Evolution of the
accelerated energy spectra over time. The simulated spectra at
t=1, 3, 5, 7, 9, and 11 minutes (solid circles) are shown along with the
upstream $\kappa=4$ distribution (black dashed curve). The hollow diamonds
represent 5 minute sector averages of $ACE$/EPAM LEMS30 and LEMS120
data taken immediately following the shock for 1998 August 26.
The vertical red line indicates the injection energy $E_{inj}$.
\label{tacc}}
\end{figure}

We first present the result for the event that occurred on 1998 August 26. 
Figure \ref{tacc} shows the accelerated energy spectra from simulations with the 
increasing of the acceleration time (colored solid circles) along with
$5$-minute sector averages (red diamonds) of $ACE$/EPAM LEMS30 and LEMS120 data 
taken immediately after the shock. The dashed curve shows the original upstream 
kappa distribution with $\kappa=4$. It is clearly seen that as the time increases 
the accelerated spectrum is gradually hardened, and after about $11$ minutes, 
the spectrum reaches the steady state which is in good agreement with the 
observation, we thus assume the acceleration time of this event $t_{acc}$ to be 
$11$ minutes. It is noted that the injection energy $E_{inj}$ for this 
event is 1.6 keV indicated by the vertical red line in Figure \ref{tacc}.

\begin{figure}
\epsscale{.8} \plotone{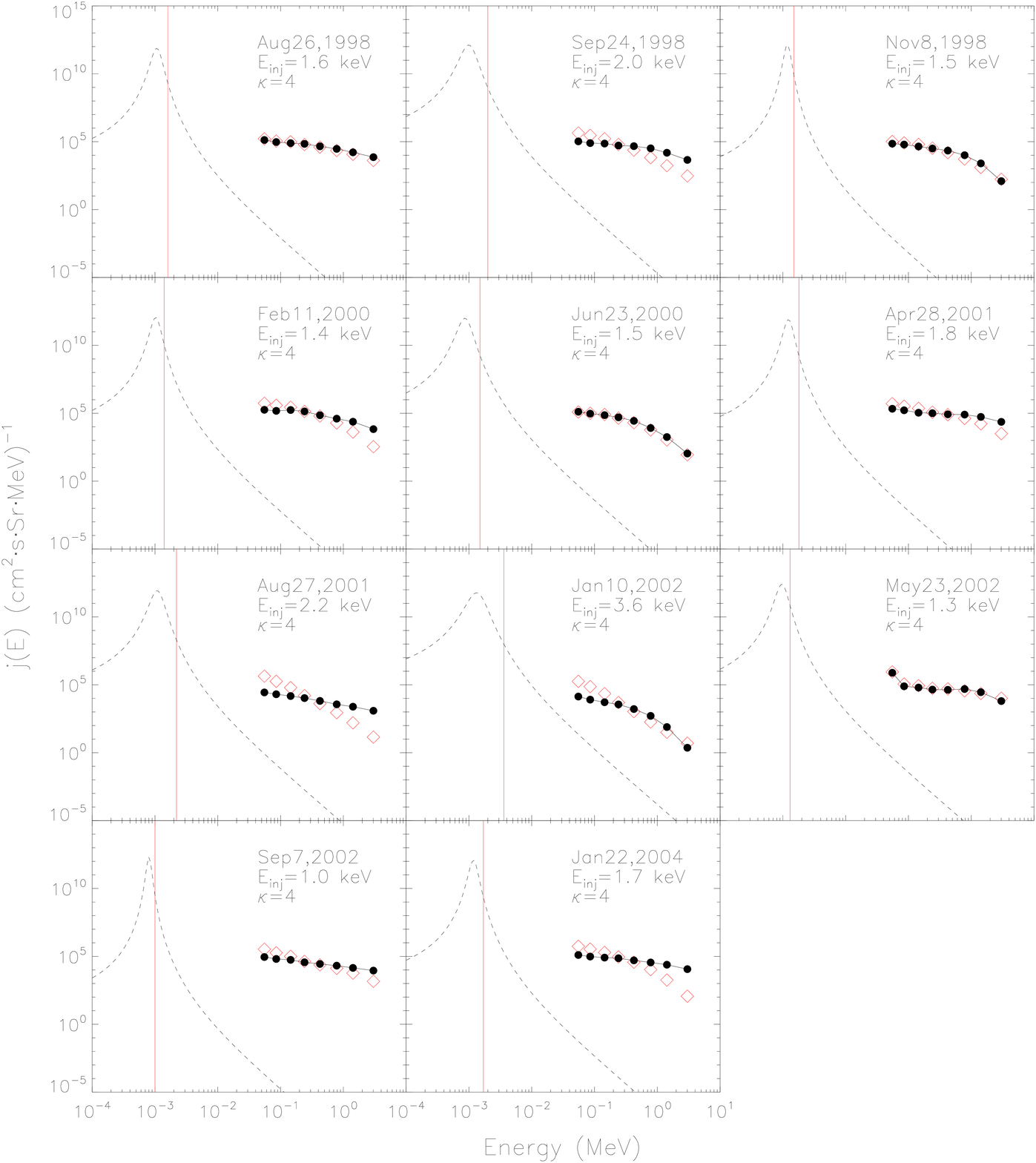} \caption{For each shock event with acceleration 
time
$t_{acc}<1$ hour, the accelerated energy spectrum at the shock front
from numerical simulations (solid circles) is shown along with
the upstream kappa distribution (dashed line). The hollow diamonds
are the EPAM observations immediately following the shock.
The injection energy is indicated by the vertical line.\label{cat1}}

\end{figure}

Similarly, we can get the acceleration time $t_{acc}$ and injection energy $E_{inj}$
for all of the quasi-perpendicular shock events. Figure \ref{cat1} shows the 
accelerated energy spectra at the shock front for all of the $11$ shock events 
with $t_{acc}\le 1$ hour. The black solid circles are the accelerated energy
spectra from simulations with EPAM observations (red diamonds) overlaid. 
For the four events occurring on 1998 August 26, 1998 November 8, 2000 June 23, 
and 2002 May 23, the accelerated spectra fit well with the observations. However, 
for the rest seven cases, the simulated energy spectra are harder than the 
observations. These results indicate that our model is very efficient in 
accelerating particles with energies from a few tens of keV to several MeV.

\begin{figure}
\epsscale{1.} \plotone{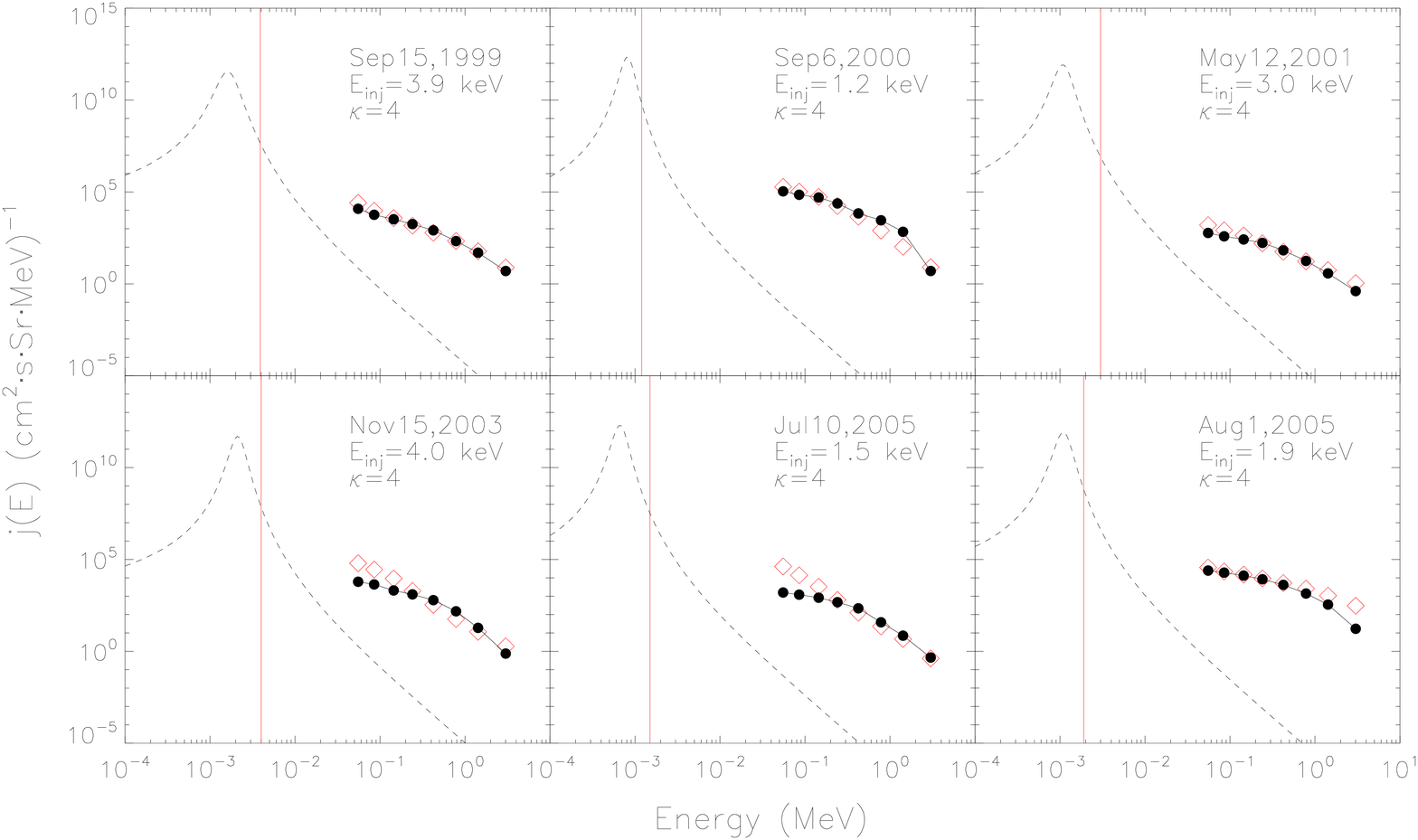} \caption{Similar as Figure \ref{cat1} except that
the acceleration time $1<t_{acc}<20$ hours.\label{cat2}}
\end{figure}

Figure \ref{cat2} is similar as Figure \ref{cat1} except that the shock events 
are with the acceleration time $t_{acc}$ ranging from $1$ to $20$ hours. 
The accelerated energy spectra from simulations for these events are in general 
fit well with the observed spectra. As can be seen, the 1999 September 15 and 
the 2001 May 12 events show the best agreement between simulations and observations. 
It is worth noting that the obviously slower upstream speeds of these events, 
which are within 130--200 km s$^{-1}$, compared to those of most shocks 
in the first subcategory are responsible for the longer acceleration time.

\begin{figure}
\epsscale{1.} \plotone{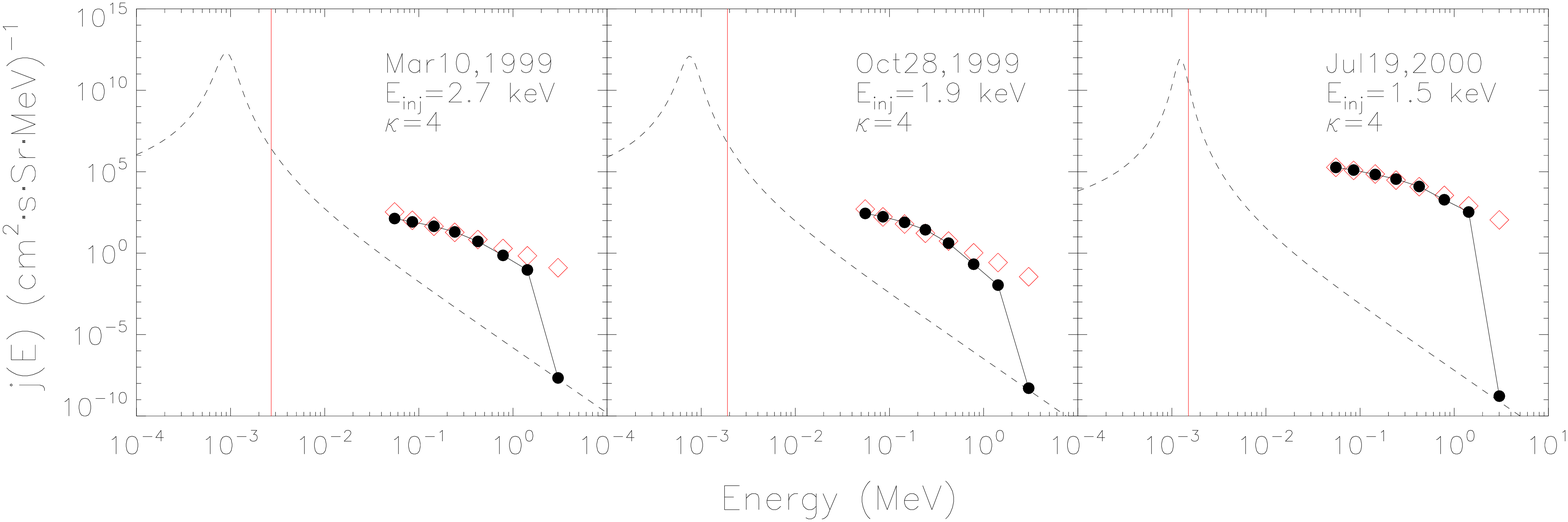} \caption{Similar as Figure \ref{cat1} except that
the acceleration time $t_{acc}>20$ hours.\label{cat3}}
\end{figure}

We also simulated a number of shock events with even slower upstream speeds which 
are about tens of kilometers per second and found that all of these shocks are 
very difficult to accelerate particles to higher energies within $20$ hours. 
In principle, simulations should be performed for the time long enough with an 
upper limit of several days considering the finite age of interplanetary shocks. 
However, we here set an upper limit of acceleration time of $20$ hours, 
i.e., a $20-$hours cutoff exactly, in the simulations to 
ensure the single shocks in our model. We list three representative 
examples in Table \ref{inputpara2} and plot the simulated results in Figure \ref{cat3}. 
From Figure \ref{cat3} we can see that all of the three shocks do not have the 
ability to accelerate particles to energies of $\sim 3$ MeV with the acceleration 
time of $20$ hours, therefore, all of their acceleration time $t_{acc}>20$ hours.

\begin{figure}
\epsscale{1.5} \plotone{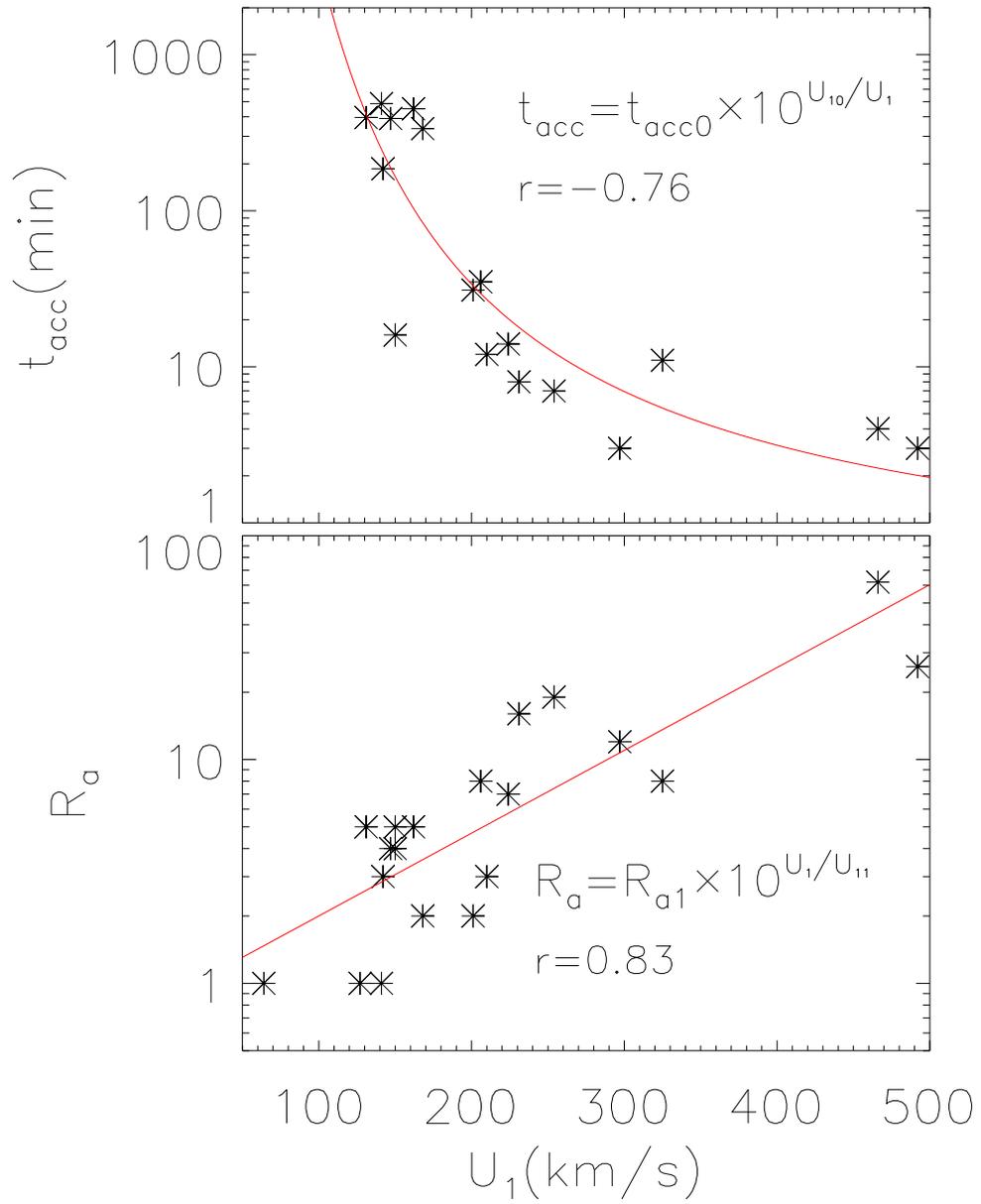} \caption{Top panel: The acceleration time 
$t_{acc}$ obtained from simulations for the
17 shocks in subcategory 1 and 2 versus the upstream speed $U_1$ shown as asterisks.
Also plotted is a fitting to the inverse proportional function 
(red curve). Bottom panel: The flux ratio $R_a$ of 47--65 keV protons as a function of
the upstream speed $U_1$ for all 20 shocks studied in this work. The red line represents
a linear fitting. \label{taccU}}
\end{figure}

For all shock events, the acceleration time $t_{acc}$ with the shock obliquities 
$\theta_{Bn}$ and upstream speeds $U_1$ are shown in Table \ref{taccEinj}. We plot 
the relationship between $t_{acc}$ and $U_1$ for the events with 
$t_{acc}<20$ hours in the top panel of Figure \ref{taccU} as asterisks.
The red curve in top panel of Figure \ref{taccU} shows the fitting to an inverse 
proportional function in log-linear space with the form 
\begin{equation}
t_{acc}=t_{acc0}\times 10^{U_{10}/U_1},
\end{equation}
where $t_{acc0}=0.291$ minutes, $U_{10}=413$ km s${}^{-1}$, 
and the correlation coefficient, $r=-0.76$. 
The fact that $r<-0.7$ indicates a negative correlation between the acceleration 
time and the upstream speed. For the shock events with $U_1$ larger than 
$200$ km s$^{-1}$, $t_{acc}$ is less than 1 hour, while for most events with 
$U_1$ in the range of 130--200 km s$^{-1}$, $t_{acc}$ is in the range of 1--20 
hours. The shocks with even lower $U_1$, such as those during the 1999 March 10 
and 1999 October 28 events, are found not capable of accelerating sufficient 
particles to high energies within $20$ hours. 

There are some individual events deviating from the general trend discussed 
above, as the observed shock events are very complicated and the acceleration time 
can be affected by many shock parameters such as the shock-normal angle, 
upstream speed, magnetic field strength, compression ratio and so on.
The shock occurred on 2001 August 27, which is classified into subcategory 1 
as shown in Figure \ref{cat1}, has $t_{acc}=16$ minutes but an upstream speed 
$U_1=150$ km s$^{-1}$ which is slower than $200$ km s$^{-1}$. The short 
acceleration time for this event may be expected to be related to other 
factors, such as the shock obliquity discussed below. This shock 
has a normal angle of $89^\circ$, which is very close to $90^\circ$ and may 
compensate for the low $U_1$ to result in $t_{acc}<$ 1 hour. On the other hand, 
there are some events in subcategory 2 with $U_1$ around $150$ km s$^{-1}$ and 
$\theta_{Bn}$ 3--9$^\circ$ deviated from $90^\circ$ in which $t_{acc}$ is about 
a couple of hours. Therefore, it is suggested that the shock geometry 
($\theta_{Bn}$) also plays an important role in determining the acceleration 
time besides the upstream speed.

To better understand the role of upstream speed in the shock acceleration, 
we examine the energetic particles intensity during the shock crossing 
for all the $20$ shock events. Here we define a quantity $R_a$ from spacecraft 
observations as the ratio of the flux of 47--65 keV at the time of the shock 
crossing to the background flux which is determined from the averaged flux of 
12 points around 2 hours preceding the shock arrival. The values of the flux 
ratio $R_a$ are shown in Column 7 of Table \ref{taccEinj}, and we plot the 
relationship between $R_a$ and $U_1$ in the bottom panel of Figure \ref{taccU} 
as asterisks. 
In addition, the red line in the bottom panel of Figure \ref{taccU} shows a 
linear fitting in log-linear space with the form
\begin{equation}
R_a=R_{a1}\times 10^{U_1/U_{11}},\label{eq:Ra}
\end{equation}
where $R_{a1}=0.853$, $U_{11}=270$ km s$^{-1}$,
and the correlation coefficient $r=0.83$. It is noted that the upstream 
speed is not the only parameter to control particle acceleration, and as the shock event 
conditions are complicated with many shock parameters varying, a relatively large 
upstream speed does not always correspond to a relatively high flux ratio. 
However, we here intend to focus on the general trend between $U_1$ and $R_a$.
The fact that the correlation coefficient is large ($r=0.83$) indicates the 
positive correlation between $R_a$ and $U_1$. Considering that the flux ratio 
$R_a$ is actually a measure of the shock ability for accelerating particles to 
high energies, the observational results with Equation (\ref{eq:Ra}) are well 
consistent with our simulation results showing a strong acceleration ability for 
the shocks with large upstream speeds, which can be explained from the 
theoretical view: assuming a fixed background magnetic field, a large upstream 
speed will yield a high convective electric field which contributes to a large 
flux at the shock crossing.

We also consider an upstream Maxwellian distribution in order for a comparison 
with the kappa distribution for each shock event. The injection energy $E_{inj}^K$ 
and $E_{inj}^M$ from kappa and Maxwellian distributions, respectively, 
are shown in Table \ref{Einj_K,M}. We find that in each shock event 
$E_{inj}^K>E_{inj}^M$. Furthermore, most of the values of $E_{inj}^K$ and $E_{inj}^M$ 
lie within the theoretical range of 1--10 keV obtained in the condition of 
$\langle b^2\rangle/B^2=0.78$ in \citet{Zank2006JGR...111..A06108Z}.
Therefore, the Maxwellian distribution can produce an accelerated distribution 
that matches the data. The higher injection energy for the kappa distribution is due to 
the fact that it is a more appropriate representation of the suprathermal population 
for quasi-perpendicular shocks.

\section{DISCUSSION AND CONCLUSIONS}

In this paper, we perform test particle simulations of particle acceleration 
associated with quasi-perpendicular interplanetary shocks with modeling magnetic 
turbulence by solving the Newton-Lorentz equation numerically with a 
backward-in-time method. We identify a set of quasi-perpendicular shocks from 
the $ACE$ shock database at 1 AU from 1998 to 2005, and use the observed solar
wind parameters to construct kappa and Maxwellian functions as the
background upstream particle distributions \citep{NeergaardParker2014ApJ...782..52P}.
By comparing the accelerated energy spectra between simulations and observations,
we find that the shocks are capable of accelerating thermal particles to high 
energies of the order of MeV with both kappa and Maxwellian upstream particle 
distributions. In addition, the injection energy and timescale of 
particle acceleration are obtained. Through examining the relationship between 
the acceleration time and the parameters such as upstream speed and shock-normal 
angle $\theta_{Bn}$, we clarify the crucial shock features that are responsible 
for efficient particle acceleration.

It it noted that our simulation results indicate that we can get accelerated energy 
spectra with the Maxwellian upstream distribution from quasi-perpendicular shocks. 
\citet{NeergaardParker2014ApJ...782..52P} also calculated accelerated 
distribution from the Maxwellian using the final kappa injection energy. 
It was shown that in many cases, the obtained accelerated distribution from the 
Maxwellian is slightly less than or comparable to that produced by the kappa 
distribution. However, in our simulations for all cases the injection energy from 
the Maxwellian distribution is less than that produced by the kappa distribution.
This difference may originate from the fact that we used test particle simulations
by solving the Newton-Lorentz equation while \citet{NeergaardParker2014ApJ...782..52P}
theoretically solved the steady-state transport equation, and we adopted a turbulence 
model in which the local background magnetic field has a component parallel to 
the shock normal so the shock is not precisely quasi-perpendicular locally. 
Although in our simulations for the quasi-perpendicular shocks the shock normal 
is almost perpendicular to large scale background magnetic field, it is usually 
less perpendicular to the local background magnetic field, since in our magnetic 
turbulence model the local averaged magnetic field deviates from the large scale 
averaged one. Therefore the injection energy in theory of perpendicular shock 
acceleration should not be a problem in our model.

In the future, we plan to study the particle acceleration at quasi-parallel 
interplanetary shocks with the similar models as in this work.


\acknowledgments
The authors thank Drs. Yang Wang, Gang Li, and Chao-Sheng Lian for 
useful discussions related to this study. 
We thank the $ACE$ EPAM, SWEPAM and MAG instrument teams, and 
the ACE Science Center for providing the ACE data. We also thank
Qiang Hu and Vasiliy Vorotnikov for help with the $ACE$ shock database.
Our work are partly supported by grants NNSFC 41574172, NNSFC 41374177, and NNSFC 
41125016.

\clearpage

\begin{table}
\caption {Summary of Parameters Used to Construct Upstream Kappa
and Maxwellian Thermal Particle Distributions.\label{inputpara1}}
\begin{tabular} {ccccccc}\hline\hline
Date of Shock && $n_1$(cm$^{-3}$) && $T_1$($^\circ$ K)  && $U_{sw}$(km s$^{-1}$) 
\\\hline
1998 Aug 26   && 4.25 && $7.17\times 10^4$  && 451 \\
1998 Sep 24   &&10.51 && $1.45\times 10^5$  && 434 \\
1998 Nov  8   && 5.39 && $3.24\times 10^4$  && 475 \\
1999 Mar 10   &&11.32 && $6.79\times 10^4$  && 414 \\
1999 Sep 15   && 3.69 && $2.44\times 10^5$  && 554 \\
1999 Oct 28   && 6.03 && $5.21\times 10^4$  && 379 \\
2000 Feb 11   && 5.78 && $6.01\times 10^4$  && 443 \\
2000 Jun 23   && 6.57 && $1.00\times 10^5$  && 405 \\
2000 Jul 19   && 3.67 && $3.76\times 10^4$  && 487 \\
2000 Sep  6   &&10.53 && $4.94\times 10^4$  && 394	\\
2001 Apr 28   && 4.26 && $6.67\times 10^4$  && 484 \\
2001 May 12   && 5.97 && $1.15\times 10^5$  && 454 \\
2001 Aug 27   && 6.16 && $1.18\times 10^5$  && 454 \\
2002 Jan 10   && 6.76 && $2.92\times 10^5$  && 501 \\
2002 May 23   &&14.87 && $7.66\times 10^4$  && 429 \\   	
2002 Sep  7   && 4.97 && $1.19\times 10^4$  && 392 \\
2003 Nov 15   && 4.19 && $1.55\times 10^5$  && 635 \\
2004 Jan 22   && 5.84 && $5.51\times 10^4$  && 475 \\
2005 Jul 10   && 9.33 && $4.76\times 10^4$  && 356 \\
2005 Aug  1   && 5.62 && $9.48\times 10^4$  && 457 \\\hline
\end{tabular}
\end{table}

\begin{table}
\caption {Shock Parameters Including the Date, Shock Obliquity, Shock 
Speed, Upstream Speed, Mach Number, Compression Ratio, Upstream Magnetic Field 
Strength,
Turbulence parameter $c$, and the Upstream Convective Electric Field. The Upstream
Background Magnetic Field is Obtained from the $320$ Seconds Averaged $ACE$/MAG 
DATA.
\label{inputpara2}}
\begin{tabular} {ccccccccc}\hline\hline
Date of Shock&$\theta_{Bn}(^\circ)$ & $V_{sh}$(km/s)&$U_1$(km/s)&$M_A$ & s & $B_0$(nT)& 
c($b/B$)
& $U_1\times B_0$($10^{-3}$N/C)\\\hline
1998 Aug 26 & 88 & 675 & 325 & 6.3 & 3.2 & 4.79  & 0.107 & 1.56\\
1998 Sep 24 & 87 & 644 & 297 & 2.9 & 3.0 & 14.67 & 0.047 & 4.35\\
1998 Nov  8 & 77 & 582 & 210 & 1.2 & 1.9 & 17.35 & 0.044 & 3.55\\
1999 Mar 10 & 78 & 522 & 127 & 3.3 & 1.5 & 5.19  & 0.047 & 0.64\\
1999 Sep 15 & 83 & 648 & 168 & 2.3 & 2.3 & 7.33  & 0.054 & 1.22\\
1999 Oct 28 & 83 & 423 &  64 & 1.2 & 1.7 & 6.84  & 0.069 & 0.43\\
2000 Feb 11 & 88 & 539 & 224 & 3.6 & 2.8 & 6.71  & 0.045 & 1.50\\
2000 Jun 23 & 85 & 483 & 201 & 3.0 & 2.5 & 8.23  & 0.055 & 1.65\\
2000 Jul 19 & 81 & 606 & 150 & 3.0 & 2.9 & 4.64  & 0.071 & 0.69\\
2000 Sep  6 & 85 & 485 & 131 & 2.3 & 2.3 & 8.73  & 0.029 & 1.14\\
2001 Apr 28 & 88 & 905 & 492 & 5.9 & 3.7 & 9.35	 & 0.089 & 4.60\\
2001 May 12 & 84 & 486 & 141 & 1.2 & 1.3 & 13.09 & 0.018 & 1.84\\
2001 Aug 27 & 89 & 486 & 150 & 2.7 & 2.8 & 6.35  & 0.085 & 0.95\\
2002 Jan 10 & 75 & 643 & 206 & 3.5 & 3.0 & 10.08 & 0.087 & 2.01\\
2002 May 23 & 84 & 834 & 466 & 4.2 & 1.7 & 16.12 & 0.045 & 7.47\\
2002 Sep  7 & 89 & 628 & 231 & 2.4 & 2.9 & 8.44	 & 0.061 & 1.95\\
2003 Nov 15 & 81 & 738 & 162 & 3.0 & 2.6 & 6.54  & 0.080 & 1.05\\
2004 Jan 22 & 89 & 738 & 254 & 4.2 & 3.7 & 7.14	 & 0.102 & 1.81\\
2005 Jul 10 & 82 & 374 & 147 & 2.0 & 1.8 & 10.64 & 0.030 & 1.55\\
2005 Aug  1 & 87 & 496 & 142 & 1.5 & 1.9 & 6.25  & 0.051 & 0.89\\\hline	
\end{tabular}
\end{table}
\begin{table}
\caption {Shock Obliquity, Upstream Speed, Acceleration Time, Injection Energy,
and the Ratio of the Intensity for the Shock Events.\label{taccEinj}}
\begin{tabular} {cccccc}\hline\hline
Date of Shock&$\theta_{Bn}$&$U_1$(km/s)&$t_{acc}$(min)&$E_{inj}$(keV)&$R_a$(47--65 keV)
\\\hline
1998 Aug 26 &88 &325 & 11	& 1.6 &  8\\		
1998 Sep 24 &87 &297 & 3	& 2.0 & 12\\
1998 Nov  8 &77 &210 & 12	& 1.5 &  3\\
1999 Mar 10 &78 &127 &$>$1200	& 2.7 &  1\\
1999 Sep 15 &83 &168 & 336	& 3.9 &  2\\
1999 Oct 28 &83 & 64 &$>$1200	& 1.9 &  1\\		
2000 Feb 11 &88 &224 & 14  	& 1.4 &  7\\
2000 Jun 23 &85 &201 & 31  	& 1.5 &  2\\
2000 Jul 19 &81 &150 & $>$1200 	& 1.5 &  4\\
2000 Sep  6 &85 &131 & 396 	& 1.2 &  5\\
2001 Apr 28 &88 &492 & 3  	& 1.8 & 26\\
2001 May 12 &84 &141 & 486   	& 3.0 &  1\\
2001 Aug 27 &89 &150 & 16  	& 2.2 &  5\\
2002 Jan 10 &75 &206 & 35  	& 3.6 &  8\\
2002 May 23 &84 &466 & 4  	& 1.3 & 62\\
2002 Sep  7 &89 &231 & 8  	& 1.0 & 16\\
2003 Nov 15 &81 &162 & 450  	& 4.0 &  5\\
2004 Jan 22 &89 &254 & 7  	& 1.7 & 19\\
2005 Jul 10 &82 &147 & 390	& 1.5 &  4\\
2005 Aug  1 &87 &142 & 186	& 1.9 &  3\\\hline

\end{tabular}
\end{table}

\begin{table}
\caption {Injection Energy for Kappa and Maxwellian Upstream Distributions 
for the Shock Events.\label{Einj_K,M}}
\begin{tabular} {ccc}\hline\hline
Date of Shock&$E^K_{inj}$(keV)&$E^M_{inj}$
(keV)\\\hline
1998 Aug 26 & 1.6 & 1.43 \\		
1998 Sep 24 & 2.0 & 1.54 \\
1998 Nov  8 & 1.5 & 1.42 \\
1999 Mar 10 & 2.7 & 1.40 \\
1999 Sep 15 & 3.9 & 2.65 \\
1999 Oct 28 & 1.9 & 1.13 \\		
2000 Feb 11 & 1.4 & 1.30 \\
2000 Jun 23 & 1.5 & 1.26 \\
2000 Jul 19 & 1.5 & 1.45 \\
2000 Sep  6 & 1.2 & 1.08 \\
2001 Apr 28 & 1.8 & 1.60 \\
2001 May 12 & 3.0 & 1.70 \\
2001 Aug 27 & 2.2 & 1.66 \\
2002 Jan 10 & 3.6 & 2.38 \\
2002 May 23 & 1.3 & 1.20 \\
2002 Sep  7 & 1.0 & 0.93 \\
2003 Nov 15 & 4.0 & 3.05 \\
2004 Jan 22 & 1.7 & 1.53 \\
2005 Jul 10 & 1.5 & 0.99 \\
2005 Aug  1 & 1.9 & 1.55 \\\hline
\end{tabular}
\end{table}


\begin{thebibliography}{33}
\bibitem[{{Axford} {et~al.}(1977){Axford}, {Leer}, \&
  {Skadron}}]{Axford1977ICRC...11..132A}
{Axford}, W.~I., {Leer}, E., \& {Skadron}, G. 1977, Proc. 15th ICRC, 11, 132
\bibitem[{{Bell}(1978)}]{Bell1978MNRAS...182..147B}
{Bell}, A.~R. 1978, \mnras, 182, 147
\bibitem[{{Bieber} {et~al.}(1996){Bieber}, {Wanner}, \&
  {Matthaeus}}]{Bieber1996JGR...101..2511B}
{Bieber}, J.~W., {Wanner}, W., \& {Matthaeus}, W.~H. 1996, \jgr, 101, 2511
\bibitem[{{Blandford} \& {Ostriker}(1978){Blandford}, \& 
{Ostriker}}]{Blandford1978ApJ...221..L29B}
{Blandford}, R.~D., \& {Ostriker}, J.~P. 1978, \apj, 221, L29
\bibitem[{{Decker} \& {Vlahos}(1986a){Decker}, \& 
{Vlahos}}]{Decker1986ApJ}
    {Decker}, R.~B., \& {Vlahos}, L. 1986a, \apj, 306, 710
\bibitem[{{Decker} \& {Vlahos}(1986b){Decker}, \& 
{Vlahos}}]{Decker1986JGR...91..13349D}
    {Decker}, R.~B., \& {Vlahos}, L. 1986b, \jgr, 91, 13349
\bibitem[{{Ellison}(1981)}]{Ellison1981GRL...8..991E}
{Ellison}, D.~C. 1981, \grl, 8, 991
\bibitem[{{Ellison} \& {Ramaty}(1985){Ellison}, \& 
{Ramaty}}]{Ellison1985ApJ...298..400E}
    {Ellison}, D.~C., \& {Ramaty}, R. 1985, \apj, 298, 400
\bibitem[{{Ellison} {et~al.}(1995){Ellison}, {Baring}, \& 
{Jones}}]{Ellison1995ApJ...453..873E}
{Ellison}, D.~C., {Baring}, M.~G., \& {Jones}, F.~C. 1995, \apj, 453, 873
\bibitem[{{Giacalone} {et~al.}(1992){Giacalone}, {Burgess}, {Schwartz}, 
\& {Ellison}}]{Giacalone1992GRL...19..433G}
{Giacalone}, J., {Burgess}, D., {Schwartz}, S.~J., 
\& {Ellison}, D.~C. 1992, \grl, 19, 433
\bibitem[{{Giacalone} \& {Jokipii}(1999){Giacalone}, 
\& {Jokipii}}]{Giacalone1999ApJ...520..204G}
    {Giacalone}, J., \& {Jokipii}, J.~R. 1999, \apj, 520, 204
\bibitem[{{Giacalone} \& {Ellison}(2000){Giacalone}, 
\& {Ellison}}]{Giacalone2000JGR...105..12541G}
    {Giacalone}, J., \& {Ellison}, D.~C. 2000, \jgr, 105, 12541
\bibitem[{{Giacalone}(2003)}]{Giacalone2003PSS...51..659G}
{Giacalone}, J. 2003, \planss, 51, 659
\bibitem[{{Giacalone}(2005)}]{Giacalone2005ApJ...628..L37G}
{Giacalone}, J. 2005, \apj, 628, L37
\bibitem[{{Giacalone} \& {K\'ota}(2006){Giacalone}, \& 
{K\'ota}}]{Giacalone2006SSR...124..277G}
{Giacalone}, J., \& {K\'ota}, J. 2006, \ssr, 124, 277
\bibitem[{{Giacalone}(2015)}]{Giacalone2015ApJ...799..80G}
{Giacalone}, J. 2015, \apj, 799, 80
\bibitem[{{Gray} {et~al.}(1996){Gray}, {Pontius}, \&
  {Matthaeus}}]{Gray1996GRL...23..965G}
{Gray}, P.~C., {Pontius}, D.~H., Jr., 
\& {Matthaeus}, W.~H. 1996, \grl, 23, 965
\bibitem[{{Heras} {et~al.}(1992){Heras}, {Sanahuja}, {Smith}, {Detman}, \&
  {Dryer}}]{Heras1992}
{Heras}, A.~M., {Sanahuja}, B., {Smith}, Z.~K., {Detman}, T., \&
{Dryer}, M.
  1992, \apj, 391, 359
  \bibitem[{{Kallenrode} \& {Wibberenz}(1997)}]{Kallenrode1997}
{Kallenrode}, M.~B., \& {Wibberenz}, G. 1997, \jgr, 102, 22311
\bibitem[{{Krymsky}(1977)}]{Krymsky1977DoSSR...234..1306K}
{Krymsky}, G.~F. 1977, Dokl. Akad. Nauk SSSR, 234, 1306   
\bibitem[{{Li} {et~al.}(2003){Li}, {Zank}, \& {Rice}}]{Li2003}
{Li}, G., {Zank}, G.~P., \& {Rice}, W.~K.~M. 2003, \jgr, 108, 1082
\bibitem[{{Matthaeus} {et~al.}(1990){Matthaeus}, {Goldstein}, \&
  {Roberts}}]{Mattaeus1990JGR...95..20673M}
{Matthaeus}, W.~H., {Goldstein}, M.~L., 
\& {Roberts}, D.~A. 1990, \jgr, 95, 20673
\bibitem[{{Neergaard Parker} \& {Zank}(2012){Neergaard Parker}, \&
  {Zank}}]{NeergaardParker2012ApJ...757..97P}
{Neergaard Parker}, L., \& {Zank}, G.~P. 2012, \apj, 757, 97
\bibitem[{{Neergaard Parker} {et~al.}(2014){Neergaard Parker}, {Zank}, \&
  {Hu}}]{NeergaardParker2014ApJ...782..52P}
{Neergaard Parker}, L., {Zank}, G.~P., \& {Hu}, Q. 2014, \apj, 782, 52
\bibitem[{Qi} {et~al.}(2017)]{QiEA17}
{Qi}, S.-Y., {Qin}, G., \& {Wang}, Y. 2017, Res. Astron. Astrophys., 17, 33
\bibitem[{{Qin}(2002)}]{Qin2002PhDT...1Q}
{Qin}, G. 2002, PhD thesis, UNIVERSITY OF DELAWARE
\bibitem[{{Qin} {et~al.}(2002a){Qin}, {Matthaeus}, \& {Bieber}}]{QinEA02GRL}
{Qin}, G., {Matthaeus}, W.~H., \& {Bieber}, J.~W. 2002a, \grl, 29, 1048
\bibitem[{{Qin} {et~al.}(2002b){Qin}, {Matthaeus}, \& {Bieber}}]{QinEA02APJ}
{Qin}, G., {Matthaeus}, W.~H., \& {Bieber}, J.~W. 2002b, \apj, 578, L117
\bibitem[{Qin} {et~al.}(2013)]{QinEA13}
{Qin}, G., {Wang}, Y., {Zhang}, M., \& {Dalla}, S. 2013, \apj, 766, 74
\bibitem[{{Quest}(1988)}]{Quest1988JGR...93..9649Q}
{Quest}, K.~B. 1988, \jgr, 93, 9649
\bibitem[{{Rice} {et~al.}(2003){Rice}, {Zank}, \&
  {Li}}]{Rice2003}
{Rice}, W.~K.~M., {Zank}, G.~P., \& {Li}, G. 2003, \jgr, 108, 1369
\bibitem[{{Scholer}(1990)}]{Scholer1990GRL...17..1821S}
{Scholer}, M. 1990, \grl, 17, 1821
\bibitem[{Wang} {et al.}(2012)]{WangEA12}
{Wang}, Y., {Qin}, G., \& {Zhang}, M. 2012, \apj, 752, 37
\bibitem[{{Zank} \& {Matthaeus}(1992){Zank}, \& {Matthaeus}}]{Zank1992JGR...97..17189Z}
    {Zank}, G.~P., \& {Matthaeus}, W.~H. 1992, \jgr, 97, 17189
\bibitem[{{Zank} {et~al.}(2000){Zank}, {Rice}, \&
  {Wu}}]{Zank2000}
{Zank}, G.~P., {Rice}, W.~K.~M., \& {Wu}, C.~C. 2000, \jgr, 105, 25079
\bibitem[{{Zank} {et~al.}(2006){Zank}, {Li}, {Florinski}, {Hu},
{Lario}, \& {Smith}}]{Zank2006JGR...111..A06108Z}
{Zank}, G.~P., {Li}, G., {Florinski}, V., {Hu}, Q., {Lario}, D., 
\& {Smith}, C.~W. 2006, \jgr, 111, A06108
\bibitem[{{Zhang} {et~al.}(2017){Zhang}, {Qin}, \& {Sun}}]{ZhangEA17}
{Zhang}, L.-H., {Qin}, G., {Sun}, P., \& Wang, H.-N. 2017, submitted to \apj,  
arXiv:1702.04647
\end{thebibliography}
\end{document}